\documentclass[aps,pra,reprint,superscriptaddress,10pt]{revtex4-2}

\usepackage{graphicx}
\usepackage{comment}
\usepackage{siunitx}
\usepackage{amsmath}
\usepackage{amssymb}
\usepackage{xcolor}

\usepackage[
  colorlinks=true,
  linkcolor=blue,
  citecolor=blue,
  urlcolor=blue
]{hyperref}

\begin{document}

\title{Overcoming intrinsic material limitations through cavity feedback}

\author{M.~Ebrahimi}
\affiliation{Department of Physics, University of Alberta, Edmonton, Alberta T6G~2E9, Canada}

\author{Y.~Huang}
\affiliation{Department of Physics, University of Alberta, Edmonton, Alberta T6G~2E9, Canada}

\author{V.A.S.V.~Bittencourt}
\affiliation{Institut de Science et d’Ingénierie Supramoléculaires (ISIS, UMR7006), Universit\'{e} de Strasbourg, 67000 Strasbourg, France}

\author{A.~Rashedi}
\affiliation{Department of Physics, University of Alberta, Edmonton, Alberta T6G~2E9, Canada}

\author{A.~Metelmann}
\affiliation{Institut de Science et d’Ingénierie Supramoléculaires (ISIS, UMR7006), Universit\'{e} de Strasbourg, 67000 Strasbourg, France} 
\affiliation{Institute for Theory of Condensed Matter and Institute for Quantum Materials and Technology, Karlsruhe Institute of Technology, 76131, Karlsruhe, Germany}

\author{J.P.~Davis}
\email[Corresponding author: ]{jdavis@ualberta.ca}
\affiliation{Department of Physics, University of Alberta, Edmonton, Alberta T6G~2E9, Canada}

\begin{abstract}

Magnons, the quanta of spin waves, have significant potential for use in modern technologies, especially when strongly coupled to another mode for read-out and control.  However, while magnons strongly interact with microwave photons via the magnetic-dipole interaction to form hybrid cavity--magnon polariton modes, the weak magnetostrictive magnon--phonon interaction, together with large polariton linewidths dominated by magnon dissipation, has so far restricted magnonic-spheres to the weak-coupling regime. The material-limited magnon dissipation rate in particular has been regarded as an unavoidable limitation in these systems. Here, we surpass this long-standing limitation by implementing an active microwave feedback loop to suppress the linewidth of cavity--magnon polaritons and strongly suppress their effective decay rate below the magnon-limited linewidth, thereby enhancing the polariton--phonon cooperativity from $C \simeq 1$ to $C \simeq 150$. As a key milestone, we achieve normal-mode splitting between a cavity--magnon polariton and a mechanical mode, providing direct evidence of three-mode hybridization among photons, magnons, and phonons. Our results establish feedback as a general route to accessing strong-coupling regimes in systems previously thought to be limited by material properties and hence open new opportunities for coherent control in hybrid quantum systems.

\end{abstract}

\maketitle

\section{Introduction}

The premise of hybrid quantum systems is to use the best properties of individual subsystems to create  new, complementary, functionalities \cite{Kurizki2015}.  For example, magnons offer the possibility of new magnetic sensing paradigms \cite{Trickle2020, Chang2026}, or nonreciprocity, enabling new circulators and isolators \cite{Li2022_Hybrid, Kim2024, Jia2026} or nonreciprocal wavelength-transducers \cite{Engelhardt2022}.  When introduced into a microwave cavity, magnons interact strongly with cavity photons forming cavity polaritons \cite{Soykal2010, Huebl2013, Tabuchi2014, Goryachev2014,Potts2020, Ebrahimi2021, SadatEbrahimi2023}, deemed cavity magnonics.  Magnons and MHz-frequency phonons, e.g., from elastic breathing modes of sub-mm spheres \cite{Zhang2016, Potts2021}, couple via a radiation-pressure-like interaction: elastic vibrations modulate the magnon resonance frequency, and the magnetization precession causes elastic deformations. This effect forms the basis of cavity magnomechanics, giving rising to the magnetostrictive version of dynamical backaction effects \cite{Zhang2016, Potts2021}. A distinctive feature of cavity magnomechanics is the triple-resonance condition, in which the phonon frequency matches the frequency difference between the hybrid cavity--magnon polariton modes, enabling selective enhancement of magnon--phonon scattering processes \cite{Zhang2016, Potts2021}. This regime has attracted significant interest and motivated proposals for a wide range of applications, including the generation of nonclassical entangled states~\cite{Li2018, Li2019, Nair2020, Li2021, Ding2021, Cheng2021}, squeezed states~\cite{Li2019_squeezed}, classical and quantum information processing~\cite{Li2020, Sarma2021, Zhao2021}, quantum correlation thermometry~\cite{Potts2020_Thermometry}, quantum non-demolition measurement~\cite{Bittencourt2025},  and the exploitation of parity--time (PT) symmetry~\cite{Ding2021, Wang2019}.

Despite these appealing features, cavity magnomechanics has so far remained firmly in the weak-coupling regime. While the photon--magnon interaction can readily reach the strong-coupling regime, the parametric magnetostrictive magnon--phonon interaction is weak in sub-mm spheres, where the photon--magnon interaction is the strongest. More importantly, the linewidth of the hybrid cavity--magnon polariton is dominated by intrinsic magnon dissipation arising from material properties, since the microwave photon dissipation rate can be tailored by engineering the microwave cavity ~\cite{Suleymanzade2020, Kudra2020, Shi2022}. Such material-limited dissipation rates cannot be overcome by increasing the drive strength alone and have therefore been widely regarded as an unavoidable barrier preventing access to the strong magnomechanical coupling regime. As a result, although hybrid cavity--magnon interactions have been studied both theoretically~\cite{Cao2019, Rao2020, Harder2021,  Ebrahimi2021, ZareRameshti2022, SadatEbrahimi2023, Gardin2023} and experimentally~\cite{Zhang2014, Potts2020,  Wolz2020, Xu2020,  Xu2021, Li2022, Girich2023, Yao2023,  Yao2025, Song2025, Wang2025, Lambert2025}, experimental investigations of magnomechanical coupling have remained limited, with all studies confined to the weak-coupling regime ~\cite{Zhang2016, Potts2021, Ebrahimi2026}.

\begin{figure*}[t]
   \centering
   \includegraphics[width=.9\linewidth]{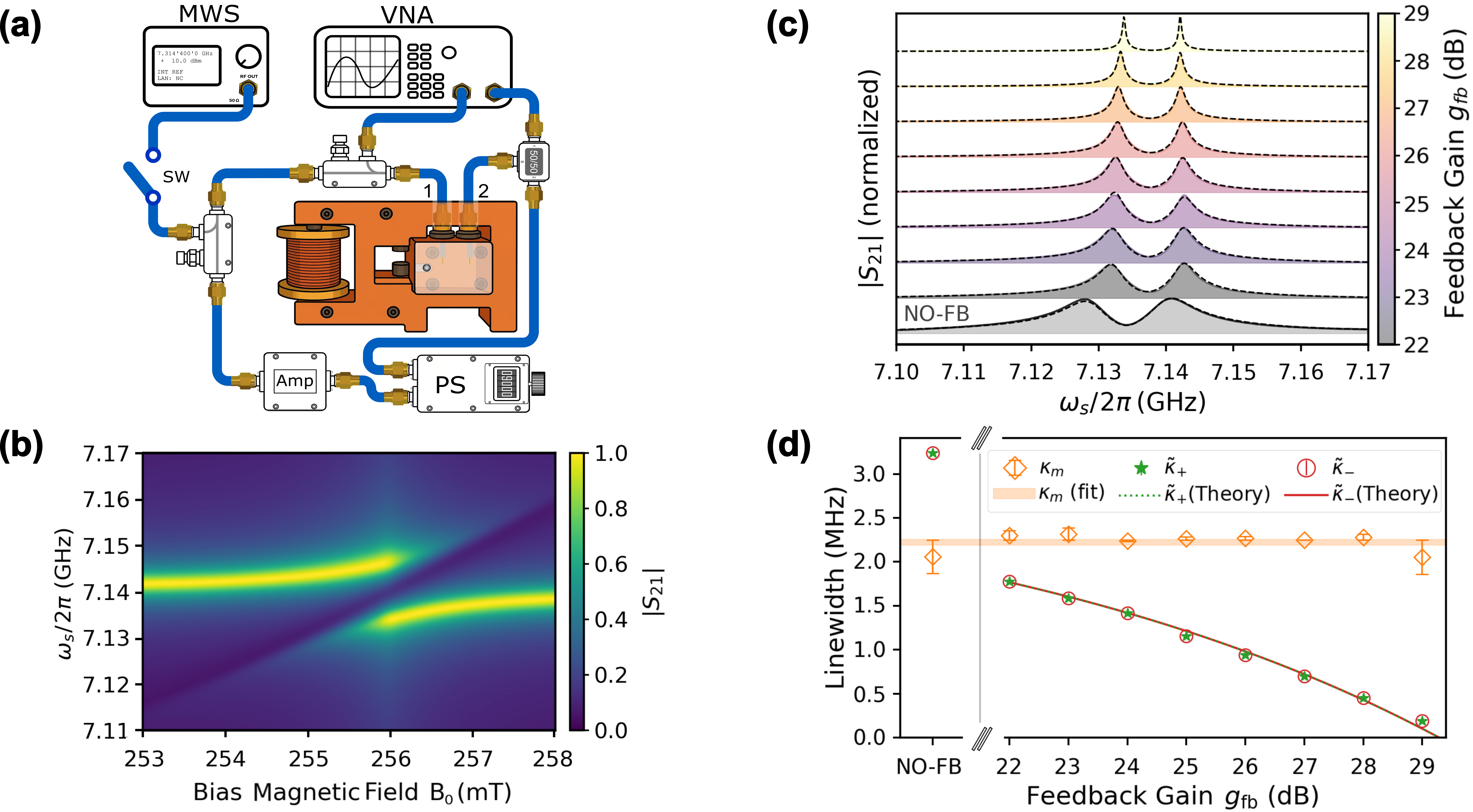}
\caption{Feedback reduces the hybridized polariton dissipation below the magnon linewidth. 
(a) Schematic of the cavity magnomechanical system and microwave feedback loop. The copper cavity is surrounded by a tunable magnet and probed by a vector network analyzer (VNA) through port~1.  A fraction of the cavity output (port 2) is directed to the VNA for measurement, while the remaining signal is routed through the feedback loop. In the feedback loop, the signal is phase shifted (PS), amplified (Amp), and subsequently fed back into the cavity through port~1. A pump tone can be added to the circuit using an independent microwave source (MWS). (b) Polariton spectra demonstrating strong coupling between magnons and cavity photons as the bias field is tuned.  (c) Transmission amplitude of the cavity--magnon polariton modes with increasing feedback gain $g_{\mathrm{fb}}$. Each trace is normalized and vertically offset for clarity. For each spectrum, the bias magnetic field was adjusted to maintain symmetric polariton modes. 
In the absence of feedback (NO-FB), the system exhibits broad hybrid modes, while increasing feedback gain progressively narrows the polariton linewidths. (d) Extracted linewidths of the upper ($\tilde{\kappa}_{+}$) and lower ($\tilde{\kappa}_{-}$) polariton modes, together with the intrinsic magnon dissipation rate $\kappa_m$, as functions of the feedback gain $g_{\mathrm{fb}}$. The feedback-induced linewidth reduction suppresses the polariton dissipation more than an order of magnitude below the magnon-limited linewidth. 
}
   \label{Figure_1}
 \end{figure*}

Feedback provides a powerful framework for controlling hybrid quantum systems by actively modifying their dynamical properties \cite{Rossi2017, Rossi2018, Zippilli2018, Serafini2020, Ebrahimi2022, Ernzer2023, Du2025}. In conventional feedback schemes, outcomes are processed and fed back to the system to engineer its effective response \cite{Wiseman2010}. In cavity opto- and electromechanical systems, feedback techniques have been employed to achieve mechanical cooling \cite{Rossi2017, Ernzer2023, Du2025}, importantly in the unresolved-sideband regime.  More recently, feedback has been applied directly to the cavity to suppress effective cavity dissipation \cite{Rossi2017, Rossi2018}, enabling optomechanical normal-mode splitting and access to the strong-coupling regime even when the system parameters lie in the weak-coupling regime \cite{Rossi2018}.  Nonetheless, feedback in optomechanical systems gives no clear advantage with respect to the cooperativity, as similar cooperativity enhancements can be obtained by applying stronger drive tones. This is not the case for hybrid cavity--magnon polaritons, and as such here we demonstrate -- both theoretically and experimentally -- the clear advantage that can be gained in this system by employing feedback to overcome intrinsic material limitations in the magnonic system.     

We experimentally implement an active microwave feedback loop to engineer the dissipation of cavity--magnon polaritons. By feeding the cavity output back into the system with tunable gain and phase, we achieve suppression of the effective polariton decay rate by more than an order of magnitude, to below the magnon dissipation rate, overcoming an intrinsic limitation of the system. As a consequence, the magnomechanical coupling rate exceeds the total dissipation rate, leading to the observation of the long-sought normal-mode splitting between a mechanical mode and a cavity--magnon polariton. 
By introducing feedback, we enhance the magnomechanical cooperativity from $C \simeq 1$ to $C \simeq 150$. 
The resulting high magnomechanical cooperativity constitutes a key prerequisite for further studies of coherent energy exchange ~\cite{Teufel2011-1, Grblacher2009, Rossi2018}, ground-state mechanical cooling ~\cite{Schliesser2008, Teufel2011-2, Rossi2017, Chan2011}, and transduction between magnonic, mechanical, and microwave degrees of freedom ~\cite{Engelhardt2022}. These results demonstrate that microwave feedback provides a general route to overcoming material-limited dissipation, enabling access to strong-coupling regimes that were previously unattainable. Our results establish a deterministic three-mode hybrid system and represent a starting point for future experimental studies in cavity magnomechanics, especially those reaching into the quantum regime ~\cite{Li2018, Li2019, Nair2020, Potts2020, Li2021, Ding2021, Cheng2021, Sarma2021, Lu2021, Chakraborty2023, Amazioug2023, Qian2024, Lu2025}.

\section{Experimental setup and feedback-enabled overcoming intrinsic magnon dissipation}

Our experiment employs a three-dimensional copper microwave cavity, allowing application of bias magnetic fields, with inner volume $30 \times 30 \times 2.5~\mathrm{mm}^3$, operated at room temperature and supporting a fundamental $\mathrm{TE}_{101}$ mode \cite{Potts2021} at $\omega_a/2\pi \approx \SI{7.1338}{\giga\hertz}$. The cavity is coupled to two external ports, enabling independent access for weak probing, coherent pump tones, and the implementation of an active microwave feedback loop. The intrinsic cavity decay rate is $\kappa_{\mathrm{int}}/2\pi \approx \SI{3.18}{\mega\hertz}$, while the total external coupling rate, with both ports connected, is $\kappa_{\mathrm{ext}}/2\pi = (\kappa_{\mathrm{ext}}^{(1)}+\kappa_{\mathrm{ext}}^{(2)})/2\pi \approx \SI{1.24}{\mega\hertz}$.

A single-crystal yttrium iron garnet (YIG) sphere with a radius of $200~\mu\mathrm{m}$ and an intrinsic magnon dissipation rate $\kappa_m/2\pi = \SI{2.20}{\mega\hertz}$ is positioned near the magnetic-field antinode of the cavity mode yielding the magnon-microwave coupling rate $g_{ma}/2\pi = \SI{5.84}{\mega\hertz}$. The orientation of the sphere is fixed by a static magnetic field applied along its easy axis. This field is generated by a pair of  permanent magnets, as shown schematically in Fig.~\ref{Figure_1}(a), and can be finely tuned using a 104-turn solenoid wound around a pure iron core and coupled via an iron yoke~\cite{Tabuchi2015}. 
The bias magnetic field enables precise tuning of the uniform magnon mode frequency $\omega_m = \gamma B_0$, where $\gamma = 28\,\mathrm{GHz/T}$ and $B_0$ is the bias  field, thereby allowing control of the cavity--magnon polariton modes, as shown in Fig.~\ref{Figure_1}(b).

To investigate feedback-induced modifications of the cavity–magnon polariton modes in the absence of mechanical excitation, the cavity is first probed through port~1 using a weak microwave signal. The output field from port~2 is divided using a power splitter: one arm is directed to a vector network analyzer (VNA) for transmission measurements, while the other arm is routed through a  phase shifter and tunable-gain microwave amplifier before being fed back into the cavity through port~1 via a directional coupler, as illustrated in Fig.~\ref{Figure_1}(a).

\begin{figure*}[t]
   \centering
   \includegraphics[width=.97\linewidth]{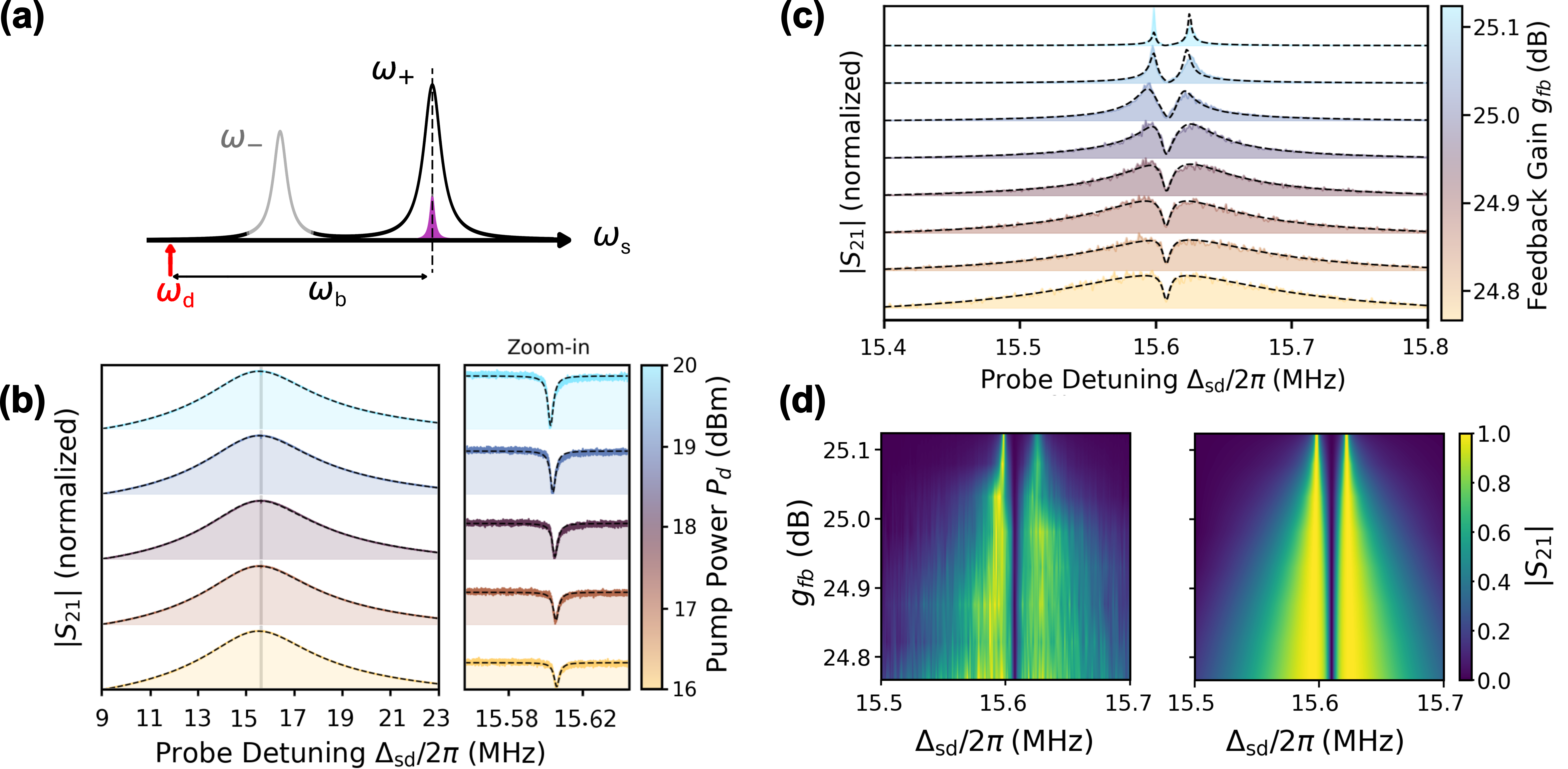}
\caption{Feedback-induced strong magnomechanical coupling. (a) 
Frequency schematic of the measurement showing the pump tone, $\omega_d$, approximately one mechanical frequency, $\omega_b$, red-detuned from the upper polariton mode, $\omega_+$. (b) Transmission amplitude, $|S_{21}|$, of the upper cavity--magnon polariton as a function of the probe detuning $\Delta_{sd}=\omega_s-\omega_d$ in the absence of feedback. As the red-detuned pump tone is increased, a  transparency peak appears, reflecting interference due to the mechanical spectrum, due to the cavity enhanced coupling. 
Despite the increasing drive power, the transparency feature remains small and no normal-mode splitting is observed. (c) When feedback gain is applied, there is a clear transparency peak that evolves into a well-defined normal mode splitting as the gain is increased (with constant pump tone power, $P_d=18\,\mathrm{dBm}$).
(d) Contour plots of the measured (left) and predicted (right) transmission as a function of $g_{\mathrm{fb}}$, highlighting the emergence and evolution of the normal-mode splitting with increasing feedback gain.}
   \label{Figure_2}
 \end{figure*}

In the absence of feedback, the microwave cavity--magnon system is governed by the standard microwave photon--magnon interaction ~\cite{Zhang2014, Tabuchi2014}. The introduction of an active microwave feedback loop alters the cavity response by effectively modifying its resonance frequency to 
\begin{equation}
\tilde{\omega}_a \equiv \omega_a+2\sqrt{\kappa_{\mathrm{ext}}^{(1)}\kappa_{\mathrm{ext}}^{(2)}}\,g_{\mathrm{fb}}\sin\phi,
\end{equation}
and its decay rate to 
\begin{equation}
\tilde{\kappa}_a\equiv \kappa_a -2\sqrt{\kappa_{\mathrm{ext}}^{(1)}\kappa_{\mathrm{ext}}^{(2)}}\,g_{\mathrm{fb}}\cos\phi,
\label{eq:cavdec}
\end{equation}
where $g_{\mathrm{fb}}$ is the feedback gain and $\phi$ is the feedback phase, both parameters that can be experimentally tuned. We present the theoretical modeling of our feedback loop in the Appendix. Correspondingly, the magnon--microwave polaritons exhibit modified frequencies $\tilde{\omega}_\pm$ and decay rates $\tilde\kappa_\pm$, obtained by diagonalizing the linear Heiseberg-Langevin equations governing the coupled mode dynamics. In the simpler case where the magnon frequency $\omega_m$ is adjusted to always be on resonance with the feedback modified cavity frequency $\tilde{\omega}_a$, the polaritons frequencies and decays are given by
\begin{align}
\tilde{\omega}_{\pm} &= \tilde{\omega}_a \pm \sqrt{g_{ma}^2 - (\tilde{\kappa}_a - \kappa_m)^2/4} \\
\tilde{\kappa}_{\pm} &= (\tilde{\kappa}_a+\kappa_m)/2,
\label{eq:polariton_eigs}
\end{align}
where we have also assumed that $g_{ma} >\vert\tilde{\kappa}_a - \kappa_m \vert /2 $, which is always the case for our experiment. 

The strong photon--magnon hybridization implies that the feedback provides direct control over the properties of the hybrid cavity--magnon polariton modes, as we can infer from Eqs.~\eqref{eq:polariton_eigs} and \eqref{eq:cavdec}: for $-\pi/2 < \phi < \pi/2$, the effective polariton linewidths $\tilde{\kappa}_\pm$ are always reduced in comparison with the no-feedback case. The optimum feedback phase is $\phi = 0$, which yields, at a fixed gain, the maximum reduction of the polariton linewidths. In this regime, the gain introduced by the feedback loop compensates for intrinsic dissipation in the system. We engineer this scenario with our setup and probe the transmission of the system, which we show in Figure~\ref{Figure_1}(c) for different feedback gains. As the feedback gain is increased, a pronounced narrowing of the polariton resonances is observed, indicating strong suppression of the effective polariton dissipation. Small deviations of the phase from zero can lead to frequency shifts at high gain values. 
To compensate for these shifts and maintain symmetric polariton splitting throughout the measurement sequence, the bias magnetic field was adjusted for each spectrum. This enables a consistent comparison of linewidth suppression across different feedback gains. 
It is noteworthy that while the symmetric polartion configuration is used in Fig.~\ref{Figure_1}(c), even narrower linewidths can be achieved for one of the polariton modes through intentional polariton linewidth asymmetry, at the expense of increased dissipation in the second mode.

The experimental data can be explained and fit using the theoretical model developed in the Appendix. From the standard input-output relations, we obtain the scattering matrix describing the reflection and transmission amplitude from the input to the output ports. The transmission amplitude from port 1 to port 2 is given by 
\begin{equation}
\label{Eq:S21CM}
S_{21}[\omega_s] = \frac{2\sqrt{\kappa_{\mathrm{ext}}^{(1)}\kappa_{\mathrm{ext}}^{(2)}}\,\tilde{\chi}_a[\omega_s]}{1+\chi_m[\omega_s]\,\tilde{\chi}_a[\omega_s]\, g_{ma}^2},
\end{equation}
where $\omega_s$ is the swept VNA probe frequency, $\chi_m[\omega_s]=\big[\kappa_m+i(\omega_m-\omega_s)\big]^{-1}$ is the magnon susceptibility, and $\tilde{\chi}_a[\omega_s]=\big[\tilde{\kappa}_a+i(\tilde{\omega}_a-\omega_s)\big]^{-1}$ shows the effective cavity susceptibility in the presence of feedback. The transmission response of the system, described by $S_{21}$, captures the resonant behavior of the system. The poles of Eq.~\eqref{Eq:S21CM} are directly related to the polariton frequency and linewidth. The absolute value $\vert S_{21}[\omega_s] \vert$, for the parameter regime of interest, has the shape of two Lorentzians with peak frequencies and linewidths corresponding to the frequencies and linewidths of the polariton modes. We then use Eq.~\eqref{Eq:S21CM} to fit our experimental data, shown in Fig.~\ref{Figure_1}(c). The model, shown in dashed lines, exhibits excellent agreement with the experimental data. From these fits, the polariton linewidths (at the maximally hybridized, symmetric polariton point) extracted using Eq.~(\ref{eq:polariton_eigs}) are reduced from approximately $\SI{3.24}{\mega\hertz}$ in the absence of feedback to as low as $\SI{174}{\kilo\hertz}$ at the highest feedback gain, thereby overcoming the intrinsic magnon dissipation rate, $\kappa_m/2\pi = \SI{2.20}{\mega\hertz}$, Fig.~\ref{Figure_1}(d).  
This demonstrates that intrinsic material-limited dissipation can be overcome through feedback.

\section{Feedback-enabled strong magnomechanical coupling}
\label{section_3}

The radiation-pressure like magnon--phonon interaction in cavity magnomechanics manifests experimentally through interference effects in the probe response. The analog of optomechanically induced transparency/absorption in cavity optomechanics \cite{Weis2010}—known as magnomechanically induced transparency/absorption (MMIT/MMIA)—arises from coherent interference between the weak probe field and the anti-Stokes (Stokes) scattering generated by a strong red- (blue-) detuned control field via mechanical motion. This interference produces a narrow transparency window or absorption dip in the probe transmission near the polariton resonance, with a linewidth set by the mechanical dissipation rate \cite{Zhang2016, Potts2021}. In this section,  we investigate this effect, which provides us a sensitive proble of the magnomechanical coupling.

The MMIT response of the upper polariton mode in cavity magnomechanics, realized using a YIG sphere with a mechanical resonance frequency $\omega_b/2\pi = \SI{15.61}{\mega\hertz}$, a single-magnon–phonon coupling rate $g_{mb}/2\pi = \SI{4.83}{\milli\hertz}$, and an intrinsic mechanical linewidth $\kappa_b/2\pi = \SI{2.06}{\kilo\hertz}$, without the feedback loop is shown in Fig.~\ref{Figure_2}(b). As the drive is tuned to the red side of the upper polariton mode (see Fig.~\ref{Figure_2}(a)), a small transparency window appears at the top of the upper polariton resonance. The intrinsically weak magnetostrictive magnon--phonon interaction, together with the large polariton linewidth prevents the formation of a wide transparency window. As the results in Fig.~\ref{Figure_2}(b) show, even increasing the drive power up to $20\,\mathrm{dBm}$ is insufficient to observe cavity--magnon polariton mechanical normal-mode splitting. This is a direct consequence of an intrinsic limitation of the system -- the decay rates prevent the magnomechanical cooperativity to be increased solely via a stronger drive.

Motivated by the experimentally demonstrated feedback-enabled polariton linewidth narrowing, we now investigate the interaction between magnon-microwave polaritons and phonons in a cavity magnomechanical system inside the feedback loop. In the following, in addition to the feedback circuit above, a drive tone is applied to port~1 using a directional coupler (Fig.~\ref{Figure_1}(a) with switch closed). This tone is tuned to the red sideband of the upper polariton mode and the drive power is fixed at $18\,\mathrm{dBm}$, for which the mechanically-induced transparency window is not clearly visible in the setup without the feedback loop. Under these otherwise identical conditions, we examine the effect of feedback on the MMIT response.

Figures~\ref{Figure_2}(c,d)  show the MMIT spectra measured at different feedback gains, fixed optimum phase $\phi\simeq0$, and under red-detuned drive. As the feedback gain is increased, the progressive suppression of the polariton linewidth enables the emergence of a clear normal-mode splitting. This splitting provides direct evidence for the onset of the strong-coupling regime between the polariton mode and the phonon mode. Because increasing the feedback gain shifts the polariton resonance frequency, the external bias field is adjusted during the measurement sequence to maintain the hybrid polariton-phonon modes near symmetric mode splitting. This procedure ensures that the observed normal-mode splitting arises from enhanced magnomechanical coupling rather than from detuning-induced asymmetry.

\begin{figure}[t]
    \centering
    \includegraphics[width=0.95\linewidth]{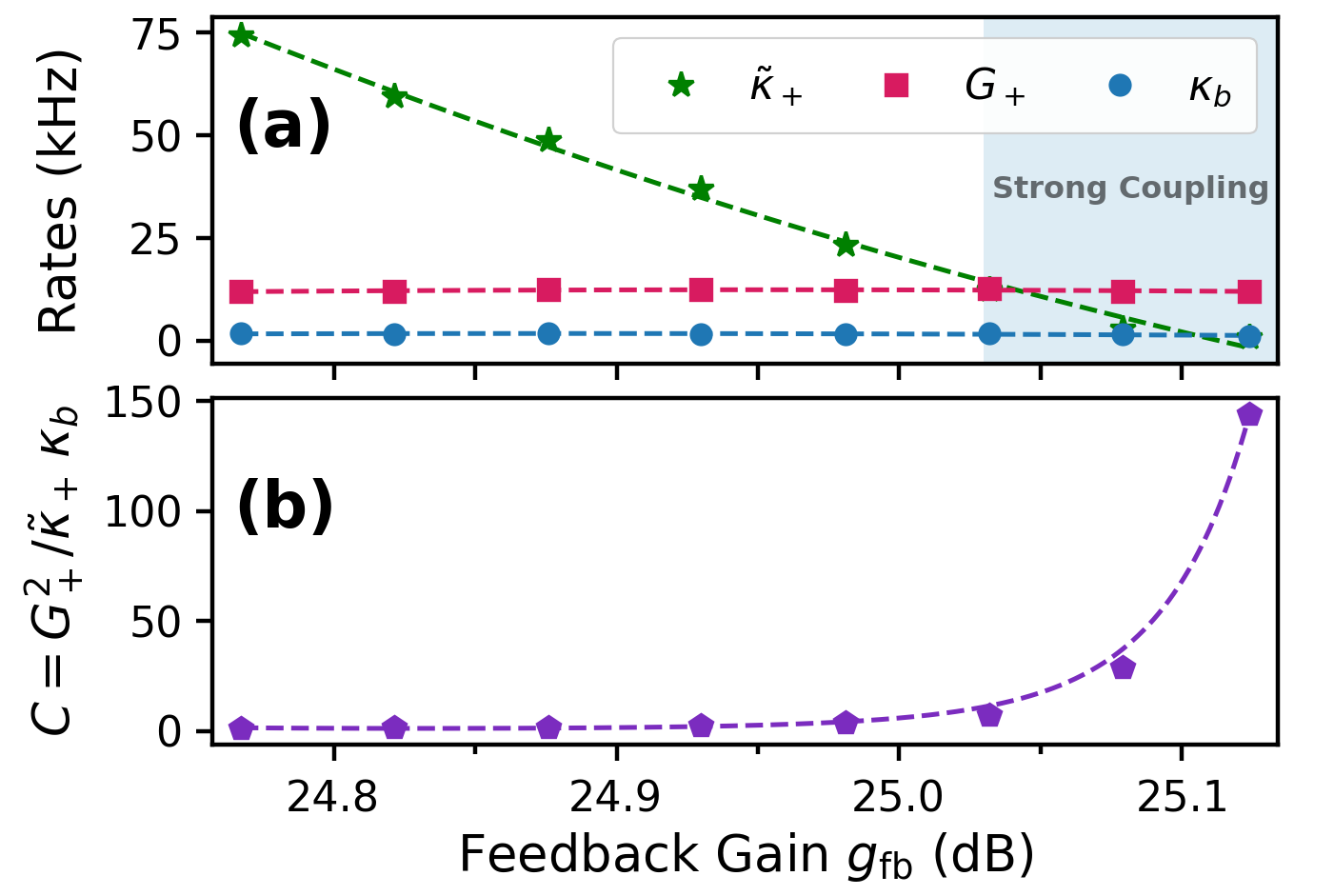}
\caption{Extracted parameters characterizing the feedback-induced strong-coupling regime.
(a) Coupling and dissipation rates extracted from fits to the transmission amplitude shown in Fig.~\ref{Figure_2}(d) as functions of the feedback gain $g_{\mathrm{fb}}$. The effective polariton--mechanical coupling strength $G_+$ and the mechanical dissipation rate $\kappa_b$ remain approximately constant, while the effective upper-polariton linewidth $\tilde{\kappa}_+$ is strongly suppressed with increasing feedback gain. The blue shaded region highlights the parameter range in which the strong-coupling condition $G_+ > \tilde{\kappa}_+, \kappa_b$ is satisfied. 
(b) Polariton--mechanical cooperativity $C = G_+^2/(\tilde{\kappa}_+ \kappa_b)$ calculated from the extracted rates as a function of the feedback gain $g_{\mathrm{fb}}$, showing a significant enhancement driven by feedback-induced linewidth suppression. 
}
\label{Mechanics_Rates}
\end{figure}

Driving the upper polariton mode in the red-detuned regime, $\omega_d = \tilde{\omega}_{+}-\omega_b$, the transmission amplitude describing the upper polariton--mechanical interaction is given by
\begin{equation}
S_{21}[\Delta_{sd}] =\frac{2\sqrt{\kappa_{+,e}\,\kappa_{\mathrm{ext}}^{(2)}}\,\tilde{\chi}_+[\Delta_{sd}] }{1+ \chi_b[\Delta_{sd}] \,\tilde{\chi}_+[\Delta_{sd}]  |G_+|^2},
\label{eq:MMIT_feedback}
\end{equation}
where $\tilde{\chi}_+[\Delta_{sd}] =\big[\tilde{\kappa}_+ + i(\omega_b-\Delta_{sd})\big]^{-1}$ and $\chi_b[\Delta_{sd}] =\big[\kappa_b + i(\omega_b-\Delta_{sd})\big]^{-1}$ denote the effective upper polariton and mechanical susceptibilities, respectively, $\Delta_{sd}=\omega_s-\omega_d$ is the probe--drive detuning, and $G_+$ represents the effective upper-polariton--phonon coupling strength. Such a transmission amplitude probes the response of the system which, just as in the case of Section II, exhibits a resonant behavior at the polariton mode frequency, including a suppression due to MMIT in the weak coupling and mode splitting in the strong coupling regime. Such behavior can be seamlessly tuned with the feedback gain parameter $g_{\mathrm{fb} }$. Theoretical spectra obtained from Eq.~(\ref{eq:MMIT_feedback}), shown as black dashed lines in Fig.~\ref{Figure_2}(c), exhibit excellent agreement with the experimental data. A small discrepancy between theory and experiment, particularly at higher feedback gain values, can be attributed to magnon Kerr nonlinearity~\cite{Shen2022} as well as magnon--phonon cross-Kerr interactions~\cite{Shen2022, Bittencourt2023}, which are neglected in our theoretical model for the sake of simplicity.

\begin{figure*}[t]
   \centering 
   \includegraphics[width=.94\linewidth]{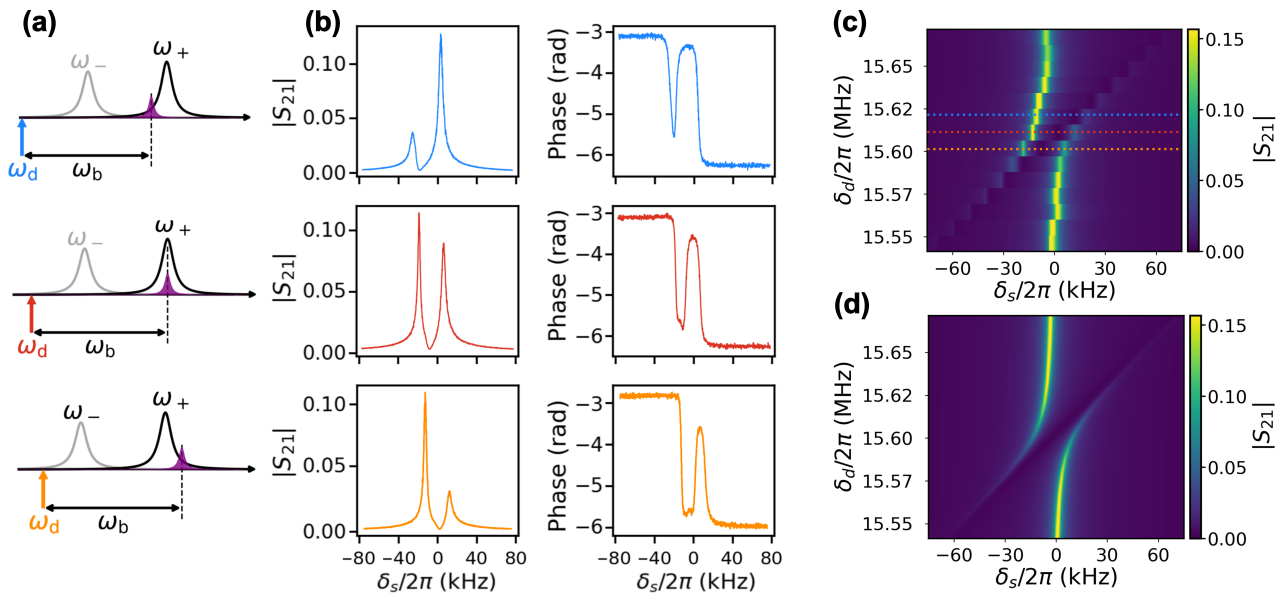}
\caption{Normal-mode (avoided level crossing) spectrum of the upper polariton--mechanical system enabled by feedback-induced linewidth suppression. (a) Schematic diagrams illustrating the relative positions of the lower and upper polariton modes ($\omega_{-}$, $\omega_{+}$), the drive/pump tone $\omega_d$, and the mechanical frequency $\omega_b$ for three representative drive frequencies (blue, orange, and red). 
(b) Corresponding line cuts of the transmission amplitude $|S_{21}|$ (left column) and phase (right column) measured at the three drive frequencies indicated in panel (a), plotted as functions of the probe detuning $\delta_s=\omega_s-\omega_0$, where $\omega_0/2\pi$ denotes the center frequency of each measurement sweep (roughly $\omega_+$). The evolution from a single resonance into two well-resolved peaks, accompanied by the characteristic phase response, reveals normal-mode splitting in the strong-coupling regime. (c) Experimental color map of the transmission amplitude $|S_{21}|$ as a function of the probe and drive detunings, $\delta_s=\omega_s-\omega_0$ and $\delta_d=\omega_0-\omega_d$, measured at a fixed probe power of $P_d=15\,\mathrm{dBm}$. A microwave feedback loop with gain $g_{\mathrm{fb}}=29.75\,\mathrm{dB}$ is applied, and the feedback phase is tuned to $\phi\simeq0$. The clear avoided crossing demonstrates coherent hybridization between the upper polariton mode and the mechanical resonance.  Colored arrows indicate the three drive frequencies corresponding to the line cuts shown in panel (b). (d) Theoretical simulation of the transmission amplitude $|S_{21}|$ plotted in the same detuning coordinates $(\delta_s,\delta_d)$, showing excellent agreement with the experimentally observed avoided-crossing spectrum.}

   \label{Figure_4}
 \end{figure*}

The fit parameters extracted from the theoretical model are summarized in Fig.~\ref{Mechanics_Rates}(a). Since the red-sideband drive power is fixed throughout the experiment at $P_d=\SI{18}{dBm}$, the cavity-enhanced polariton--phonon coupling strength $G_+$, remains approximately constant as the feedback gain is varied, as does the mechanical dissipation rate $\kappa_b$. On the other hand, increasing the feedback gain leads to a substantial reduction of the upper polariton dissipation rate $\tilde{\kappa}_+$. At sufficiently high feedback gain, the system enters the strong-coupling regime, satisfying the condition $G_+ > \tilde{\kappa}_+, \kappa_b$, as indicated by the shaded area. The transition from weak to strong polariton-phonon coupling is more clearly resolved in this way than in the more gradual transition from a transparency dip to  normal-mode splitting in Figs.~\ref{Figure_2}(c,d).

Finally, the polariton–mechanical cooperativity, $C = G_+^2 / \tilde{\kappa}_+ \kappa_b$, which quantifies the competition between coherent coupling and dissipation, is shown in Fig.~\ref{Mechanics_Rates}(b). While at low feedback gain the cooperativity is of order unity ($C \simeq 1$), increasing the feedback gain enables cooperativities as large as $C \simeq 150$. This large cooperativity establishes the cavity magnomechanical system as a promising platform for coherent quantum operations, including mechanical cooling and microwave/magnon–mechanical quantum transduction ~\cite{Schliesser2008, Grblacher2009, Teufel2011-1, Teufel2011-2, Mahboob2012, Aspelmeyer2014, Galland2014, Rossi2017, Rakhubovsky2017, Rossi2018}.

Having established the feedback-induced strong-coupling regime through linewidth suppression and enhanced cooperativity, we now present direct spectroscopic evidence of coherent hybridization between the upper polariton mode and the mechanical resonance. Figure~\ref{Figure_4} shows the transmission amplitude of a weak probe with frequency $\omega_s$ with a strong drive at a frequency $\omega_d$ red-detuned with respect to the upper polariton. The double peaked structure indicates polariton--phonon hybridization. 
As the drive frequency is swept across the red-detuned regime of the upper polariton resonance, a clear avoided crossing emerges, which is a hallmark of coherent mode hybridization and cannot be explained by interference or linewidth narrowing alone. At the red-sideband condition, the transmission transforms from a single resonance into two well-resolved peaks accompanied by a characteristic dispersive phase response, providing unambiguous evidence of normal-mode splitting arising from coherent polariton--mechanical coupling.

\section{Conclusion}

In this work, we have experimentally demonstrated that active microwave feedback can overcome intrinsic material dissipation in cavity magnomechanics, the main limiting factor in practical applications. By implementing a tunable feedback loop that modifies the effective dissipation of cavity--magnon polariton modes, we achieve a substantial suppression of the polariton linewidth to below the magnon-limited linewidth. This linewidth engineering enables the effective polariton--phonon coupling strength to exceed the total system decay rates, thereby allowing the system to enter the magnomechanical strong-coupling regime.

As a direct consequence, we observe clear normal-mode splitting between a cavity--magnon polariton and a mechanical mode, providing conclusive spectroscopic evidence of coherent hybridization among photons, magnons, and phonons. Control measurements performed under identical drive conditions but without feedback confirm that increasing the drive power alone is insufficient to reach this regime, demonstrating that feedback-enabled linewidth suppression is the key mechanism enabling strong magnomechanical coupling.

Beyond the implementation presented here, our results establish feedback-enabled linewidth engineering as a general strategy for accessing strongly coupled regimes in hybrid magnonic platforms that are otherwise limited by intrinsic magnetic losses and weak single-magnon magnomechanical coupling rates. This approach opens new opportunities for the realization of coherent information transfer ~\cite{Sarma2021, Li2021, Rakhubovsky2017}, quantum-limited control of mechanical motion ~\cite{Schliesser2008, Grblacher2009, Teufel2011-1, Teufel2011-2, Mahboob2012, Aspelmeyer2014, Galland2014, Rossi2017, Rakhubovsky2017, Rossi2018}, microwave--magnon--mechanical transduction ~\cite{Lau2020, Sarma2021}, feedback-controlled non-Hermitian and exceptional-point physics in hybrid quantum systems~\cite{Wang2019, Ding2021, Lambert2025}, and generation of entanglement between two remote mechanical resonators ~\cite{Li2021}. For such implementations in the quantum regime, it will be important to benchmark the noise introduced in the system by the feedback loop, which is known to limit, for example, the cooling performance of feedback-schemes in optomechanics \cite{Schafermeier2016, Massi_2018, Mika2025}. Even with such added noise, feedback cooling and squeezing schemes in optics and microwave-based systems are known to enable quantum control \cite{Massi_2018}, which we also anticipate to be the case in this system.

\begin{acknowledgments}
The authors acknowledge that the land on which this work was performed is on Treaty Six Territory, the traditional lands of many First Nations, Métis, and Inuit in Alberta. They acknowledge the support from the Natural Sciences and Engineering Research Council, Canada (Grant Nos.~RGPIN-2022-03078,  ALLRP 592535-23, ALLRP 568609-21); Alberta Innovates (Grant No.~212200780); the National Research Council (Grant QSP-017-1); and the University of Alberta Department of Physics through the Advah Bhatia Fellowship. A.~Metelmann and V.A.S.V.~Bittencourt acknowledge funds from the Agence nationale de la recherche, France (Grant A25R321A/2025DRI00103 ``FeEnQCaMM''). V.A.S.V.~Bittencourt thanks C.A. Potts (University of Calgary) for useful discussions.
\end{acknowledgments}


\section{Appendix A: Theoretical background}
\label{Appendix A}

Magnons and microwaves couple via a magnetic-dipole dipole interaction, and are typically close in frequency, allowing for a coherent exchange of energy. Microwave cavities can be engineered to exhibit well defined and separated frequency, such that we can consider the interaction between a magnon mode, with annihilation operator $\hat{m}$ and frequency $\omega_m$, and a single cavity mode, described with the annihilation operator $\hat{a}$ and frequency $\omega_a$. The Hamiltonian of the cavity--magnon system is given by
\begin{align}
\frac{\hat{H} }{\hbar} &=
\omega_a \hat a^\dagger \hat a
+ \omega_m \hat m^\dagger \hat m
+ g_{ma}\left(\hat m \hat a^\dagger + \hat a \hat m^\dagger\right)
\nonumber\\
&\quad
+ i\sqrt{2\kappa_{\mathrm{ext}}^{(1)}}\,\varepsilon_s
\left(\hat a^\dagger e^{-i\omega_s t}
- \hat a e^{i\omega_s t}\right),
\label{eq:Hamiltonian}
\end{align}
where the last term describes a coherent monochromatic drive with amplitude $\varepsilon_s$ applied via an external port that couples to the microwave mode with a rate $\kappa_{\mathrm{ext}}^{(1)}$. The magnon-microwave coupling rate $g_{ma}$ depends on the system architecture \cite{Bittencourt2023}. 

In our experiment, the microwave resonator is a 3D cavity, and the magnet hosting the magnons is a yttrium iron-garnet sphere with a diameter much smaller than the wave-length of the fundamental mode of the cavity. In this case, the magnon-microwave coupling depends on the position of the sphere inside the cavity and its volume. Systems like the one used in the experiment can routinely achieve strong magnon-microwave coupling \cite{Zhang2014}, which results in the formation of magnon-microwave polaritons \cite{Zhang2014, ZareRameshti2022}.

The quantum Langevin equations (QLEs) describing the system dynamics are
\begin{subequations}\label{eq:QLEs_noFB}
\begin{align}
\dot{\hat a} &=
\left(-i\omega_a-\kappa_a\right)\hat a
- i g_{ma}\hat m
+ \sqrt{2\kappa_a}\,\hat{\mathcal{A}}_{\mathrm{in}}
\nonumber\\
&\quad
+ \sqrt{2\kappa_{\mathrm{ext}}^{(1)}}\,\varepsilon_s e^{-i\omega_s t},
\label{eq:QLE_a_noFB}\\
\dot{\hat m} &=
\left(-i\omega_m-\kappa_m\right)\hat m
- i g_{ma}\hat a
+ \sqrt{2\kappa_m}\,\hat{m}_{\mathrm{in}},
\label{eq:QLE_m_noFB}
\end{align}
\end{subequations}
where $\kappa_a$ and $\kappa_m$ denote the total decay rates of the cavity and magnon modes, respectively, $\kappa_{\mathrm{ext}}^{(1)}$ is the external coupling rate through cavity port~1. The noise operator $\hat{m}_{\mathrm{in}}$ describes vacuum and thermal noise acting on the magnon mode, while $\hat{\mathcal{A}}_{\mathrm{in}}$ includes both intrinsic cavity noise, and noise fed to the cavity via its input ports.

Throughout these appendices, we will indicate the coherent part of the fields as, e.g. $a = \langle \hat{a} \rangle$. We further consider white noise exclusively, such that $\langle \hat{m}_{\rm{in}} \rangle = \langle \hat{\mathcal{A}}_{\mathrm{in}} \rangle = 0$.

\subsection{Microwave feedback}

We now introduce a feedback loop acting on the cavity mode. As illustrated schematically in Fig.~\ref{Figure_1}(a), the output field from one port of the cavity (port~2) is amplified, phase shifted, and then fed back into the cavity through port~1, which is also used to apply the probe field. The coherent amplitude of the input field at port~1 is therefore given by
\begin{align}
a_{\mathrm{in}}^{(1)}(t)
= \varepsilon_p + g_{fb}\,a_{\mathrm{out}}^{(2)}(t-\tau_{fb}),
\label{eq:ain_delay}
\end{align}
where $g_{fb}$ is the feedback gain and $\tau_{fb}$ is the feedback delay time. The feedback also adds contributions to the noise term $\hat{\mathcal{A}}_{\rm{in}}$ appearing in Eq.~\eqref{eq:QLE_m_noFB}. We will indicate the modified noise contributions by $\hat{\tilde{\mathcal{A} }}_{\rm{in}}$.

When the delay time is much shorter than both the cavity lifetime $1/\kappa_a$ and the characteristic interaction time $1/g_{ma}$, its effect can be absorbed into an effective phase shift, which allows us to write the coherent part of the input as
\begin{align}
a_{\mathrm{in}}^{(1)}(t)
= \varepsilon_s + g_{fb}\,a_{\mathrm{out}}^{(2)}(t)e^{-i\phi},
\label{eq:ain_phase}
\end{align}
where
\begin{equation}
\phi \equiv \theta_{fb} + (\omega_a-\omega_p)\tau_{fb}
\end{equation}
is the total feedback phase, with $\theta_{fb}$ controlled externally. Taking the feedback into account and using standard input--output theory \cite{Clerk2010}, the output field from port~2, with external coupling $\kappa_{\mathrm{ext}}^{(2)}$, is given by
\begin{equation}
\hat a_{\mathrm{out}}^{(2)}(t)
=
\sqrt{2\kappa_{\mathrm{ext}}^{(2)}}\,\hat a(t)
+
\hat a_{\mathrm{in}}^{(2)}(t).
\end{equation}
We emphasize that in our experiment, no coherent drive is applied via port~2, hence $\hat a_{\mathrm{in}}^{(2)}$ is composed only by thermal and vacuum noise.

The cavity quantum Langevin equation is therefore modified to
\begin{align}
\dot{\hat a}
&=
\left(-i\omega_a-\kappa_a\right)\hat a
- i g_{ma}\hat m
\nonumber\\
&\quad
+ \sqrt{2\kappa_{\mathrm{ext}}^{(1)}}
\left(
\varepsilon_s
+
\sqrt{2\kappa_{\mathrm{ext}}^{(2)}}\,g_{fb}\,\hat a\,e^{-i\phi}
\right)
e^{-i\omega_s t} \\
&\quad + \sqrt{2 \kappa_a} \hat{\tilde{\mathcal{A}}}_{\rm{in}},
\label{eq:QLE_FB}
\end{align}

Equation~\eqref{eq:QLE_FB} explicitly shows that the feedback loop introduces an effective self-interaction of the cavity field through coherent extraction, processing, and reinjection of the intracavity field.

\subsection{Effective cavity parameters}

The feedback-induced self-interaction can be absorbed into renormalized cavity parameters. We can thus rewrite  Eq.~\eqref{eq:QLE_FB} for the coherent amplitude of the cavity as
\begin{align}
\dot{ a} &=
\left(-i\tilde{\omega}_{\mathrm{a}}-\tilde{\kappa}_{\mathrm{a}}\right) a
- i g_{ma} m
+ \sqrt{2\kappa_{\mathrm{ext}}^{(1)}}\,\varepsilon_s e^{-i\omega_s t}.
\label{eq:QLE_eff}
\end{align}
The effective cavity frequency and decay rate are given by
\begin{align}
\tilde{\omega}_a &\equiv \omega_a
+ 2\sqrt{\kappa_{\mathrm{ext}}^{(1)}\kappa_{\mathrm{ext}}^{(2)}}\,g_{fb}\sin\phi,
\label{eq:omega_eff}\\
\tilde{\kappa}_a &\equiv \kappa_a
- 2\sqrt{\kappa_{\mathrm{ext}}^{(1)}\kappa_{\mathrm{ext}}^{(2)}}\,g_{fb}\cos\phi.
\label{eq:kappa_eff}
\end{align}
We emphasize that the feedback loop acts exclusively on the cavity mode, leaving the intrinsic magnon parameters $\omega_m$ and $\kappa_m$ unchanged. By tuning the feedback gain and phase, the effective cavity frequency and decay rate can be continuously engineered, enabling controlled enhancement or suppression of cavity dissipation.

\subsection{Feedback-controlled polariton modes}

Because the cavity and magnon modes are strongly coupled, it is convenient to describe the system in terms of hybrid cavity--magnon polaritons. The QLEs for the magnon and cavity coherent amplitudes in the presence of feedback read
\begin{subequations}\label{eq:QLEs_FB}
\begin{align}
\dot{ a} &=
\left(-i\tilde{\omega}_a-\tilde{\kappa}_a\right) a
- i g_{ma} m
+ \sqrt{2\kappa_{\mathrm{ext}}^{(1)}}\,\varepsilon_s e^{-i\omega_s t},
\\
\dot{ m} &=
\left(-i\omega_m-\kappa_m\right) m
- i g_{ma} a.
\end{align}
\end{subequations}
The complex eigenfrequencies of the hybrid polariton modes are given by
\begin{align}
\tilde{\omega}_{\pm}-i\tilde{\kappa}_{\pm}
&=
\frac{\tilde{\omega}_a+\omega_m}{2}
- i\frac{\tilde{\kappa}_a+\kappa_m}{2}
\nonumber\\
&\quad
\pm
\sqrt{
g_{ma}^2
+
\left(
\frac{\tilde{\omega}_a-\omega_m}{2}
- i\frac{\tilde{\kappa}_a-\kappa_m}{2}
\right)^2
}.
\label{eq:polariton_eigs_Appendix}
\end{align}
Equation~\eqref{eq:polariton_eigs_Appendix} shows that, in general, the polariton linewidths, $\tilde{\kappa}_{\pm}$, are governed by the feedback-modified cavity parameters. For analytical transparency, we consider the resonant case $\tilde{\omega}_{a}=\omega_m$, where the linewidths reduce to
\begin{align}
\tilde{\kappa}_{\pm}
=
\frac{\tilde{\kappa}_a+\kappa_m}{2}
=
\frac{\kappa_a
- 2\sqrt{\kappa_{\mathrm{ext}}^{(1)}\kappa_{\mathrm{ext}}^{(2)}}\,g_{fb}\cos\phi
+ \kappa_m}{2}.
\label{eq:kappa_pm}
\end{align}
Equation~\eqref{eq:kappa_pm} demonstrates that, although intrinsic magnon dissipation typically dominates cavity magnonic systems and limits the polariton lifetime, appropriately engineered microwave feedback allows direct control over the effective cavity dissipation.
By tuning the feedback gain and phase, the cavity mode can be driven into a regime of reduced or even negative effective dissipation, effectively behaving as an amplifying mode.
When hybridized with the lossy magnon mode, this feedback-induced amplification compensates the magnon dissipation, resulting in a pronounced suppression of the polariton linewidth.
In the ideal limit, and within the stability bounds of the feedback loop, the polariton linewidth can be made arbitrarily small.

To obtain the transmission amplitude given in Eq.~\eqref{Eq:S21CM}, we use the standard input-output relations
\begin{equation}
\hat{a}_{\rm{out}}^{(i)}(t) = \hat{a}_{\rm{in}}^{(i)}(t) - \sqrt{2 \kappa_{\rm{ext}}^{(i)}} \hat{a}(t), \quad(i = 1,2).
\end{equation}
We should notice that in the above equation $\hat{a}_{\rm{in}}^{(1)}$ is to be taken as the out-of-loop input field, i.e. $a_{\rm{in}}^{(1)}(t) = \varepsilon_{s}e^{-i \omega_s t}$.

Working in frequency domain via the Fourier transform (for $o = a,m$)
\begin{equation}
\hat{o}(t) =\int d\omega \hat{o}[\omega] e^{-i \omega t},
\end{equation}
we can solve the linear Heisenberg-Langevin equation for the coherent part of the fields, which gives us the relation between outputs and inputs
\begin{equation}
a^{(i)}_{\rm{out}}[\omega] = \sum_{j=1}^2S_{ij}[\omega] a^{(j)}_{\rm{in}}[\omega], 
\end{equation}
where $S$ is the scattering matrix of the system. Since only port~1 is coherently driven $a^{(2)}_{\rm{in}}[\omega_s] =0$ and $a^{(1)}_{\rm{in}}[\omega] =\varepsilon_s \delta[\omega - \omega_s]$. Therefore $a^{(2)}_{\rm{out}}[\omega] = S_{21}[\omega] a^{(1)}_{\rm{in}}[\omega] = S_{21}[\omega] \varepsilon_s \delta[\omega - \omega_s] $. The transmission amplitude of a coherent tone with frequency $\omega_s$ is thus $S_{21}[\omega_s]$, as given in Eq.~\eqref{Eq:S21CM}.

\subsection{Magnomechanical interaction and polariton--phonon coupling}

The magnon excitation inside the YIG sphere induces a time-dependent magnetization, which in turn deforms the sphere through magnetostrictive forces. Conversely, mechanical vibrations modify the magnetization, resulting in an intrinsic coupling between magnon and phonon modes  \cite{Zhang2016, Potts2021}. Owing to the large frequency mismatch between magnons (GHz) and phonons (MHz), this interaction is naturally weak. To compensate for this mismatch and enhance the effective interaction, a strong coherent microwave drive is applied.

The magnetostrictive interaction is described by a radiation-pressure-like dispersive Hamiltonian \cite{Zhang2016},
\begin{equation}
\hat{H}_{mb}/\hbar
=
g_{mb}\,\hat{m}^{\dagger}\hat{m}\,(\hat{b}+\hat{b}^{\dagger}),
\label{eq:H_mb}
\end{equation}
where $g_{mb}$ is the intrinsic magnomechanical coupling strength and $\hat{b}$ is the annihilation operator of the mechanical mode with resonance frequency $\omega_b$. The single-magnon magnomechanical coupling $g_{mb}$ depends on the sphere size and on magneto-elastic coefficients that are material-specific parameters \cite{Ballestero2020, Bittencourt2023}.

The system is conveniently described in terms of hybrid cavity--magnon polariton modes $\hat{A}_{\pm}$, defined as
$\hat{A}_+ = \hat{a}\cos\psi + \hat{m}\sin\psi$ and
$\hat{A}_- = -\hat{a}\sin\psi + \hat{m}\cos\psi$,
with the mixing angle $\psi$ given by
$\tan 2\psi = 2g_{ma}/(\omega_m - \tilde{\omega}_a)$.
In the presence of a coherent drive applied to the cavity at frequency $\omega_d$ with complex amplitude $\varepsilon_d$, linearization of the Hamiltonian about the steady-state polariton amplitudes yields an effective polariton--phonon interaction with enhanced coupling strengths $G_{\pm}$, proportional to the intracavity polariton population \cite{Zhang2016}.

Working in a frame rotating at the drive frequency $\omega_d$, the effective Hamiltonian of the driven cavity magnomechanical system is
\begin{align}
\frac{\hat{H}}{\hbar} &=
(\tilde{\omega}_{+}-\omega_d)\,\hat{A}_+^{\dagger}\hat{A}_+
+ (\tilde{\omega}_{-}-\omega_d)\,\hat{A}_-^{\dagger}\hat{A}_-
+ \omega_b\,\hat{b}^{\dagger}\hat{b}
\nonumber\\
&\quad
+ \left(G_+\,\hat{A}_+^{\dagger} + G_+^{*}\,\hat{A}_+\right)
(\hat{b}+\hat{b}^{\dagger})
\nonumber\\
&\quad
+ \left(G_-\,\hat{A}_-^{\dagger} + G_-^{*}\,\hat{A}_-\right)
(\hat{b}+\hat{b}^{\dagger})
\nonumber\\
&\quad
+ \sqrt{2\kappa_{+,\mathrm{ext}}^{(1)}}\,\varepsilon_s
\left(
\hat{A}_+ e^{i(\omega_s-\omega_d)t}
+ \hat{A}_+^{\dagger} e^{-i(\omega_s-\omega_d)t}
\right)
\nonumber\\
&\quad
+ \sqrt{2\kappa_{-,\mathrm{ext}}^{(1)}}\,\varepsilon_s
\left(
\hat{A}_- e^{i(\omega_s-\omega_d)t}
+ \hat{A}_-^{\dagger} e^{-i(\omega_s-\omega_d)t}
\right).
\label{eq:H_polariton_mb}
\end{align}
where 
\begin{eqnarray}
G_{+} &=& g_{mb}\left(\sin^2\psi\,\tilde{A}_{+,\mathrm{ss}} + \sin\psi\cos\psi\,\tilde{A}_{-,\mathrm{ss}}\right),\\
G_{-} &=& g_{mb}\left(\cos^2\psi\,\tilde{A}_{-,\mathrm{ss}} + \sin\psi\cos\psi\,\tilde{A}_{+,\mathrm{ss}}\right).
\end{eqnarray}
and the steady state excitation in the simple single mode regime is 
\begin{align}
\tilde{A}_{\pm,\mathrm{ss}} &=
\frac{\sqrt{2\tilde{\kappa}_{\pm,\mathrm{ext}}^{(1)}}\,\varepsilon_d}
{\tilde{\kappa}_{\pm}+i(\tilde{\omega}_{\pm}-\omega_d)}, \\
\tilde{\kappa}_{\pm,\mathrm{ext}}^{(1)} &=
\frac{1\pm \cos 2\psi}{2}\kappa_{\mathrm{ext}}^{(1)} .
\end{align}

The strong coupling regime of polariton--phonon interaction enabled by microwave feedback manifests experimentally through interference effects in the probe response, which is studied both with and without feedback in Sec.~\ref{section_3}.

The transmission amplitude of the system $S_{21}$ is obtained using the same procedure outlined in Appendix~C. The frequency at which the transmission amplitude in Eq.~\eqref{eq:MMIT_feedback} is written is the difference between the weak probe drive frequency and the strong pump tone, as the Hamiltonian~\eqref{eq:H_polariton_mb} is written in a frame rotating at the pump frequency $\omega_d$.

\begin{figure}[t]
    \centering
    \includegraphics[width=0.9\linewidth]{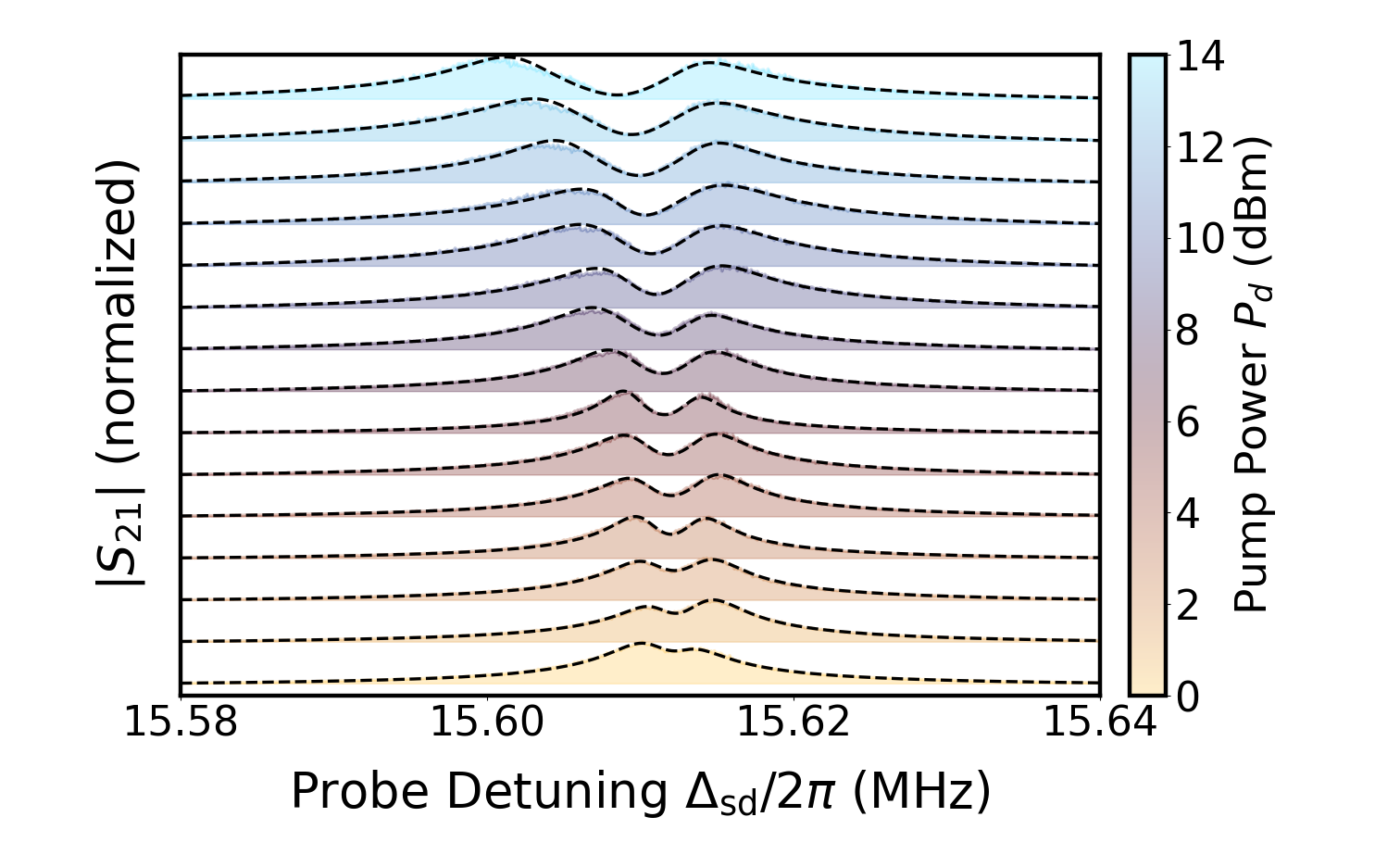}
\caption{
Transmission spectra of the cavity–magnon system measured inside the microwave feedback loop as a function of increasing drive power (bottom to top), shown for direct comparison with the no-feedback measurements in Fig.~\ref{Figure_2}(c). The feedback gain is fixed at $g_{\mathrm{fb}} = 27~\mathrm{dB}$ and the phase is tuned to $\phi\simeq 0$, corresponding to the regime of maximal polariton linewidth suppression. In contrast to the no-feedback configuration, where increasing drive power alone does not produce resolvable normal-mode splitting due to the large intrinsic polariton linewidth, the feedback-enabled reduction of the linewidth allows the drive-enhanced magnomechanical interaction to overcome the dissipation rates, leading to a clear emergence of normal-mode splitting and entry into the strong-coupling regime. Dashed lines represent fits to the theoretical model.
}    
    \label{Figure_5}
\end{figure}

\subsection{Effect of drive power with and without feedback}

To further clarify the role of drive power in the magnomechanical interaction, we investigate the transmission response of the system as a function of drive power both in the absence and in the presence of the microwave feedback loop. The measurements without feedback are presented in the main text, Fig.~\ref{Figure_2}(c), where increasing the drive power alone enhances the effective polariton--phonon coupling but does not produce resolvable normal-mode splitting because the intrinsic polariton linewidth remains large.

Figure~\ref{Figure_5} shows the corresponding measurements performed under otherwise identical conditions with the feedback loop enabled. During these measurements, the feedback gain is fixed at $g_{\mathrm{fb}} = 27\,\mathrm{dB}$ and the feedback phase is tuned to $\phi \simeq 0$, corresponding to the regime of maximal polariton linewidth suppression. In this configuration, the feedback-induced reduction of the polariton linewidth allows the drive-enhanced magnomechanical interaction to overcome the dissipation rates, leading to the clear emergence of normal-mode splitting.

These results demonstrate that increasing the drive power alone is insufficient to reach the strong magnomechanical coupling regime, and that feedback-enabled linewidth suppression is the key mechanism enabling coherent polariton--mechanical hybridization.

\end{document}